\documentclass[aps,prl,preprint,nopacs,superscriptaddress]{revtex4}

\usepackage{graphicx}
\usepackage{verbatim}
\usepackage{mathrsfs}
\usepackage{gensymb}

\usepackage{amsmath,amsfonts,amssymb}
\usepackage{graphicx}
\def\3{2.8in}    %used for figure widths
\def\2{2.5in}
\def\4{3.0in}\def \beq {\begin{equation}}
\def \eeq {\end{equation}}
\usepackage[paperwidth=8.5in,paperheight=11in,centering,hmargin=2cm,vmargin=2.5cm]{geometry} % fix margins to be regular all around
\allowdisplaybreaks

\begin{document}

\title{Experimental Observation of Two Massless Dirac-Fermion Gases in Graphene-Topological Insulator Heterostructure }

\author{Guang~Bian}
\affiliation {Laboratory for Topological Quantum Matter and Spectroscopy (B7), Department of Physics, Princeton University, Princeton, New Jersey 08544, USA}

\author{Ting-Fung~Chung}
\affiliation {Department of Physics and Astronomy, Purdue University, West Lafayette, Indiana 47907, USA}
\affiliation {Birck Nanotechnology Center, Purdue University, West Lafayette, Indiana 47907, USA}

\author{Chang~Liu}
\affiliation {Laboratory for Topological Quantum Matter and Spectroscopy (B7), Department of Physics, Princeton University, Princeton, New Jersey 08544, USA}
\affiliation {Department of Physics, South University of Science and Technology of China, Shenzhen, Guangdong 518055, China}

\author{Chaoyu~Chen}
\affiliation {Synchrotron SOLEIL, L'Orme des Merisiers, Saint-Aubin - BP48, 91192GIF-sur-YVETTE CEDEX, France}

\author{Tay-Rong~Chang}
\affiliation {Department of Physics, National Tsing Hua University, Hsinchu 30013, Taiwan}
\affiliation {Laboratory for Topological Quantum Matter and Spectroscopy (B7), Department of Physics, Princeton University, Princeton, New Jersey 08544, USA}

\author{Tailung~Wu}
\affiliation {Department of Physics and Astronomy, Purdue University, West Lafayette, Indiana 47907, USA}
\affiliation {Birck Nanotechnology Center, Purdue University, West Lafayette, Indiana 47907, USA}

\author{Ilya~Belopolski}
\affiliation {Laboratory for Topological Quantum Matter and Spectroscopy (B7), Department of Physics, Princeton University, Princeton, New Jersey 08544, USA}

\author{Hao~Zheng}
\affiliation {Laboratory for Topological Quantum Matter and Spectroscopy (B7), Department of Physics, Princeton University, Princeton, New Jersey 08544, USA}

\author{Su-Yang~Xu}
\affiliation {Laboratory for Topological Quantum Matter and Spectroscopy (B7), Department of Physics, Princeton University, Princeton, New Jersey 08544, USA}

\author{Daniel S. Sanchez}
\affiliation {Laboratory for Topological Quantum Matter and Spectroscopy (B7), Department of Physics, Princeton University, Princeton, New Jersey 08544, USA}

\author{Nasser Alidoust}
\affiliation {Laboratory for Topological Quantum Matter and Spectroscopy (B7), Department of Physics, Princeton University, Princeton, New Jersey 08544, USA}

\author{Jonathan Pierce}
\affiliation {Center for Solid State Energetics, RTI International, Research Triangle Park, North Carolina 27709, USA}

\author{Bryson Quilliams}
\affiliation {Center for Solid State Energetics, RTI International, Research Triangle Park, North Carolina 27709, USA}

\author{Philip P. Barletta}
\affiliation {Center for Solid State Energetics, RTI International, Research Triangle Park, North Carolina 27709, USA}

\author{Stephane~Lorcy}
\affiliation {Synchrotron SOLEIL, L'Orme des Merisiers, Saint-Aubin - BP48, 91192GIF-sur-YVETTE CEDEX, France}

\author{Jos$\text{\'{e}}$ Avila}
\affiliation {Synchrotron SOLEIL, L'Orme des Merisiers, Saint-Aubin - BP48, 91192GIF-sur-YVETTE CEDEX, France}

\author{Guoqing~Chang}
\affiliation {Centre for Advanced 2D Materials and Graphene Research Centre National University of Singapore, 6 Science Drive 2, Singapore 117546}
\affiliation {Department of Physics, National University of Singapore, 2 Science Drive 3, Singapore 117542}

\author{Hsin~Lin}
\affiliation {Centre for Advanced 2D Materials and Graphene Research Centre National University of Singapore, 6 Science Drive 2, Singapore 117546}
\affiliation {Department of Physics, National University of Singapore, 2 Science Drive 3, Singapore 117542}

\author{Horng-Tay Jeng}
\affiliation {Department of Physics, National Tsing Hua University, Hsinchu 30013, Taiwan}
\affiliation {Institute of Physics, Academia Sinica, Taipei 11529, Taiwan}

\author{Maria-Carmen Asensio\footnote{maria-carmen.asensio@synchrotron-soleil.fr}}
\affiliation {Synchrotron SOLEIL, L'Orme des Merisiers, Saint-Aubin - BP48, 91192GIF-sur-YVETTE CEDEX, France}

\author{Yong P. Chen\footnote{yongchen@purdue.edu}}
\affiliation {Department of Physics and Astronomy, Purdue University, West Lafayette, Indiana 47907, USA}
\affiliation {School of Electrical and Computer Engineering, Purdue University, West Lafayette, Indiana 47907, USA}
\affiliation {Birck Nanotechnology Center, Purdue University, West Lafayette, Indiana 47907, USA}

\author{M. Zahid Hasan\footnote{mzhasan@princeton.edu}}
\affiliation {Laboratory for Topological Quantum Matter and Spectroscopy (B7), Department of Physics, Princeton University, Princeton, New Jersey 08544, USA}


\date{\today}

\begin{abstract}
Graphene and topological insulators (TI) possess two-dimensional (2D) Dirac fermions with distinct physical properties. Integrating these two Dirac materials in a single device creates interesting opportunities for exploring new physics of interacting massless Dirac fermions. Here we report on a practical route to experimental fabrication of graphene-Sb$_2$Te$_3$ heterostructure. The graphene-TI heterostructures are prepared by using a dry transfer of chemical-vapor-deposition grown graphene film. ARPES measurements confirm the coexistence of topological surface states of Sb$_2$Te$_3$ and Dirac $\pi$ bands of graphene, and identify the twist angle in the graphene-TI heterostructure. The results suggest a potential tunable electronic platform in which two different Dirac low-energy states dominate the transport behavior.

\end{abstract}

\maketitle

Dirac materials such as graphene and topological insulators (TI) have been in the spotlight attracting global research interest in the past decade, because they not only provide unprecedented opportunities to explore fundamental physics in condensed matters but also open a new route for device innovations thanks to the unique electronic properties of those materials.\cite{NS_review, RMP, Zhang_RMP, graphene_RMP, Geim, Novoselov, Kim, Zhou, Rotenberg} In both graphene and TIs, the low-energy excitations can be described as a two dimensional (2D) electron gas with linear dispersion akin to massless Dirac fermions. There is a key difference between the two systems: in graphene the intrinsic spin-orbit coupling (SOC) is negligibly small and the chirality of the Dirac state is associated with the sublattice degrees freedom, the so-called pseudospin; on the other hand, in TIs the topological bulk band ordering is caused by the strong SOC and gives rise to the helical spin-momentum locked feature of the Dirac surface states. There have emerged many interesting proposals of combining the two Dirac materials in heterostructures, taking advantage of the dramatic difference in spin texture of their Dirac states.\cite{Diaz, Rossi, Jhi, Niu, Zhang}  It has been predicted that the SOC of graphene can be greatly enhanced by proximity to the strongly spin-orbit coupled TI surface. As a consequence, the epitaxial graphene can develop a spin-orbit gap as large as 20 meV, making it a Kane-Mele quantum spin-Hall insulator.\cite{Jhi, Kane}  Thus the major interest in this type of heterojunctions is the potential for engineering electronic structure of graphene by proximity effect. Despite several experimental efforts to grow TI films on graphene substrate\cite{Xue, Jiang, Dang} the experimental fabrication of a graphene layer on top of TI surface remains a tremendous technical challenge, not to mention the characterization of the electronic band structure of such graphene-TI junctions. The difficulty lies in the fact graphene can not be directly grown \textit{in situ} on the surface of TIs such as Bi$_2$Se$_3$ and Sb$_2$Te$_3$. In this work we report on an experimental route to preparing graphene-Sb$_2$Te$_3$ heterostructures in laboratory. We also perform angle-resolved photoemission (ARPES) measurements on the lab-prepared graphene-TI heterostrcuture and demonstrate the coexistence of two Dirac modes of different types. The results suggest a potential tunable electronic platform for studing the interaction between two different Dirac low-energy states.

The micro- and nano-ARPES experiments were carried out at the ANTARES beamline at the SOLEIL synchrotron \cite{Avila_1, Avila_2, Avila_3, ChenC} having respective lateral resolutions of $\sim$ 90 $\mathrm{\mu}$m and  $\sim$ 120 nm. The nanoARPES microscope is equipped with two Fresnel zone plates (FZP) to focalize the beam and an order selection aperture to eliminate higher diffraction orders. The sample was mounted on a nano-positioning stage, which was placed at the coincident focus point of the electron analyzer and the FZP. The conventional and nano-resolved photoelectron spectra were obtained using a hemispherical analyzer Scienta R4000 with 5 meV and 0.005 $\textrm{\AA}^{-1}$, energy and momentum resolutions, respectively. The Scienta detector is aligned along the $\mathrm{\Gamma}$-$\mathrm{K}$ direction of graphene. All photoemission measurements at ANTARES were done at a temperature of 100 K. The samples were degassed $\textit{in situ}$ at 200 $^\circ$C for 8 hours before ARPES measurements. The degassing was proved to be critical for acquiring good ARPES spectra. 

The first-principles calculations were based on the generalized gradient approximation (GGA)\cite{Perdew} using the full-potential projected augmented wave method\cite{Blochl} as implemented in the VASP package.\cite{Kress}  The electronic structure of bulk Sb$_2$Te$_3$ were calculated using a 15$\times$15$\times$6 Monkhorst-Pack $k$-mesh over the Brillouin zone (BZ) with the spin-orbit coupling included self-consistently. The surface states were calculated from the surface Green's function of the semi-infinite system.\cite{HJZhang} A 30$\times$30$\times$1 Monkhorst-Pack $k$-mesh was used in the calculation for graphene.

We choose Sb$_2$Te$_3$ as the TI substrate because the lattice constant of the (111) surface of Sb$_2$Te$_3$ ($a$ = 4.26 $\text{\AA}$) matches nearly perfectly a $\sqrt{3}\times\sqrt{3}$ super cell of graphene ($a$ = 2.46 $\text{\AA}$). Therefore the heterostructure in the perfect alignment can be described by a commensurate $\sqrt{3}\times\sqrt{3}$ stacking pattern. Graphene/Sb$_2$Te$_3$ heterostructures were prepared using a dry transfer of chemical vapor deposition (CVD) grown graphene film \cite{Petrone}. Graphene film was grown on Cu foils at ambient pressure which has been reported elsewhere \cite{Yu, He, Chung}.  Briefly, the growth was at 1065~$^\circ$C with 20~sccm methane (500~ppm in Ar) and 200~sccm forming gas (5\% H$_2$ in Ar) for 60-75 minutes. Prior to the growth, clean Cu foil was annealed for 120 minutes in the forming gas environment. The Sb$_2$Te$_3$ thin film was grown (at RTI International) on an undoped GaAs(100) substrate by metalorganic chemical vapor deposition (MOCVD) process. To transfer as-grown graphene film, a PMMA layer was spin-coated onto one side of the Cu foil, and the graphene on the opposite side of the Cu foil was removed by oxygen plasma. Next, a piece of wafer dicing tape (Blue Med Tack Squares from Semiconductor Equipment Corp.) with a small window cut in its center was adhered to the graphene covered with PMMA. The exposed Cu was then dissolved in ammonium persulfate solution ($\sim$2.5 g in 50 ml DI water), obtaining a suspended PMMA/graphene membrane. After Cu etching, the sample was further cleaned by DI water and SC2 baths (HCl:H$_2$O$_2$:H$_2$O=1:1:50 ml). The sample is subsequently rinsed by isopropanol and blown dry with N$_2$ gas. The typical domain size of graphene is about 20 $\mu$m. Finally, the suspended sample was transferred onto Sb$_2$Te$_3$ substrate, and followed by dissolving the PMMA layer in acetone. Figure~1(a) shows an optical micrograph of the surface of graphene/Sb$_2$Te$_3$ (thickness of Sb$_2$Te$_3$ = 1360 nm). 

A Horiba Jobin Yvon XploRA confocal Raman system with a solid-state laser operating at a wavelength of 532 nm and a spectral resolution of 1.1 cm$^{-1}$ was used for the Raman spectroscopy measurement. Power of the laser was maintained below 1 mW to minimize local heating effect. The laser spot size was ~0.6 $\mu$m in diameter with a 100$\times$ objective (NA = 0.90) and all measurements were performed in an air ambient environment and at room temperature. As shown in Fig.~1(b), there are two high-frequency Raman modes at ~1584 cm$^{-1}$ and ~2676 cm$^{-1}$ in the Raman spectra, corresponding to the G and 2D modes of graphene \cite{Malard}. We also observe a series of characteristic Raman modes at ~73 cm$^{-1}$ ($A_{1g}^{1}$), ~113 cm$^{-1}$ ($E_{g}^{2}$) and ~166 cm$^{-1}$ ($A_{1g}^{2}$) for Sb$_2$Te$_3$ \cite{Richter}. Two extra peaks could be assoicated with surface oxidation of the Sb$_2$Te$_3$ thin film. The measured Raman characteristics confirm the success in fabricating graphene/Sb$_2$Te$_3$ heterostructures.

The lattice structure of graphene/Sb$_2$Te$_3$ with a perfect angle match is plotted in Fig.~1(c).  In this case the surface unit cell of the heterostructure is same as that of Sb$_2$Te$_3$. However, in reality, there is no epitaxial relationship between CVD graphene and Sb$_2$Te$_3$ and  being van~der~Waals coupled they can be oriented in any angles. Figure~1(d) shows the photoemssion spectrum taken with 700 eV photons. The characteristic core levels of C, Sb, and Te atoms are observed, indicating the correct chemical composition of the samples. We also notice in the spectrum there exist several extra peaks which can be attributed to the contamination introduced during the $\textit{ex situ}$ assembly of the heterostructure.

The Brillouin zone corresponding the stacking geometry in Fig.~1(c) is shown in Fig.~2(a). The $K$ point of graphene (where the Dirac point of graphene is located) coincide with the $\bar{\Gamma}$ point of Sb$_2$Te$_3$ in the second zone. Under this ideal situation the Dirac cones of graphene and Sb$_2$Te$_3$ overlap in momentum space, and we can expect strong interaction between the Dirac states. In the nonideal case which there exists an angle mismatch, the Dirac cones of two materials separate in $k$ space and the interaction between them is weaker. Therefore we have a tunable system (depending on the twist angle) for studying the correlation between two Dirac low-energy modes and the resulting novel transport behaviors.  First-principles band structure of Sb$_2$Te$_3$ superimposeed with that of graphene (green solid lines) is shown in Fig.~2(b). We can see the Dirac cone of graphene indeed overlap with the topological surface states of Sb$_2$Te$_3$. Figure~2(c) presents ARPES mapping of a Sb$_2$Te$_3$ film employed as the substrate in our experiment (without $\textit{in situ}$ cleavage). The mapping at the Fermi level clearly resolve the Dirac point of the topological surface states and six lobes of bulk valence band as marked by arrows, indicating the good surface quality of the MOCVD-grown Sb$_2$Te$_3$ film despite a brief exposure to the air (which is inevitable during the transfer of graphene). A comparison of the ARPES spectrum of Sb$_2$Te$_3$ with the calculated band structure is given in Fig.~2(d). We find a good consistency between the experimental and theoretical results: the Fermi level cuts the topological surface bands at the Dirac point and also grazes the top of the bulk valence band.

ARPES results taken from our graphene/Sb$_2$Te$_3$ sample is shown in Fig.~3(a). Comparing with the result from the bare Sb$_2$Te$_3$ surface, there are two additional bright spots in the ARPES mapping of the heterostructure, as indicated by the arrows. They are from the $\pi$ band of graphene. The two spots indicate there are two major domains with different twist angles in the CVD graphene layer. The twisting geometry of the sample is illustrated in Figs.~3(b) and 3(c). According to the ARPES spectra, the distance between the Dirac band of graphene and the zone center is equal to the momentum of Sb$_2$Te$_3$ surface states in the second zone, which confirms the lattice constant match of the two Dirac materials. The twist angles of two graphene domains with respect to Sb$_2$Te$_3$ are 21$^\circ$ and 20$^\circ$, respectively.  To further show the energy dispersion of Dirac fermions in this heterostructure, we plot ARPES spectral cuts at different azimuthal angles (as marked by the dashed lines in the panel on the right hand side) in Fig.~4. At angles of 22$^\circ$, 82$^\circ$, and 142$^\circ$, we observed the Dirac surface states of Sb$_2$Te$_3$. We note that the quality of the spectrum is worse than that reported in Fig.~2, which can be attributed to the coverage of graphene as well as the contamination introduced during the transfer of graphene. At 43$^\circ$ and 103$^\circ$ we saw the linear Dirac $\pi$ band from one graphene domain. The $\pi$ band of the other graphene domain was observed at 62$^\circ$ and 122$^\circ$. Through fitting the Fermi velocity of CVD graphene is found to be (1.01$\pm$0.12)~$\times$~10$^{8}$~cm$\cdot$s$^{-1}$. In the spectral cuts only one branch of the graphene $\pi$ band shows up, see Fig.~4. It is due to a well-known selection rule on photoexcitation matrix elements \cite{Rotenberg, Liu}.  The ARRES cuts confirm the twisting geometry schematic shown in Fig.~3(c) and the coexistence of two different Dirac modes in our graphene/Sb$_2$Te$_3$ samples, which is the key conclusion of the this work.

In this work, we explored a practical approach to fabricating graphene/TI heterostructrure, a promising platform for studying interacting Dirac fermions. Our ARPES experiment demonstrated the coexistence of the Dirac low-energy states of graphene and Sb$_2$Te$_3$ in our lab-prepare sample. A twist angle between the epitaxial graphene layer and the TI substrate was identified by our photoemission measurement. This twist angle provides a tunable degree of freedom for manipulating the electronic properties of the heterojunction system. Our work presented a first step towards an ideal system of two interacting Dirac fermions with a tunable interaction strength. Further improvements on the sample preparation are desired: first, reduce the contamination introduced during the transfer of graphene;  second, increase the domain size of graphene; and third, a precise way to control the twist angle. Those efforts will refine the graphene/TI system in such a way that could eventually enable the observation of the proximity effect on the graphene $\pi$ band and potentially the quantum spin-Hall effect in the original Kane-Mele model of graphene.

\section{Acknowledgements}
This work is supported by  the US National Science Foundation (NSF) under Grant No. NSF-DMR-1006492 (MZH). Work at Princeton, Purdue and RTI are also supported by DARPA MESO program (grant N66001-11-1-4107). TFC and YPC also acknowledge support of NSF DMR (grant 0847638) on graphene research. TFC acknowledges financial support from Purdue University through Bilsland Dissertation Fellowship. TRC acknowledges visiting scientist support from Princeton University. The Synchrotron SOLEIL is supported by the Centre National de la Recherche Scientifique (CNRS) and the Commissariat $\grave{\textrm{a}}$ l$'$Energie Atomique at aux Energies Alternatives (CEA), France.

\newpage

\begin{figure}
\centering
\includegraphics[width=16cm]{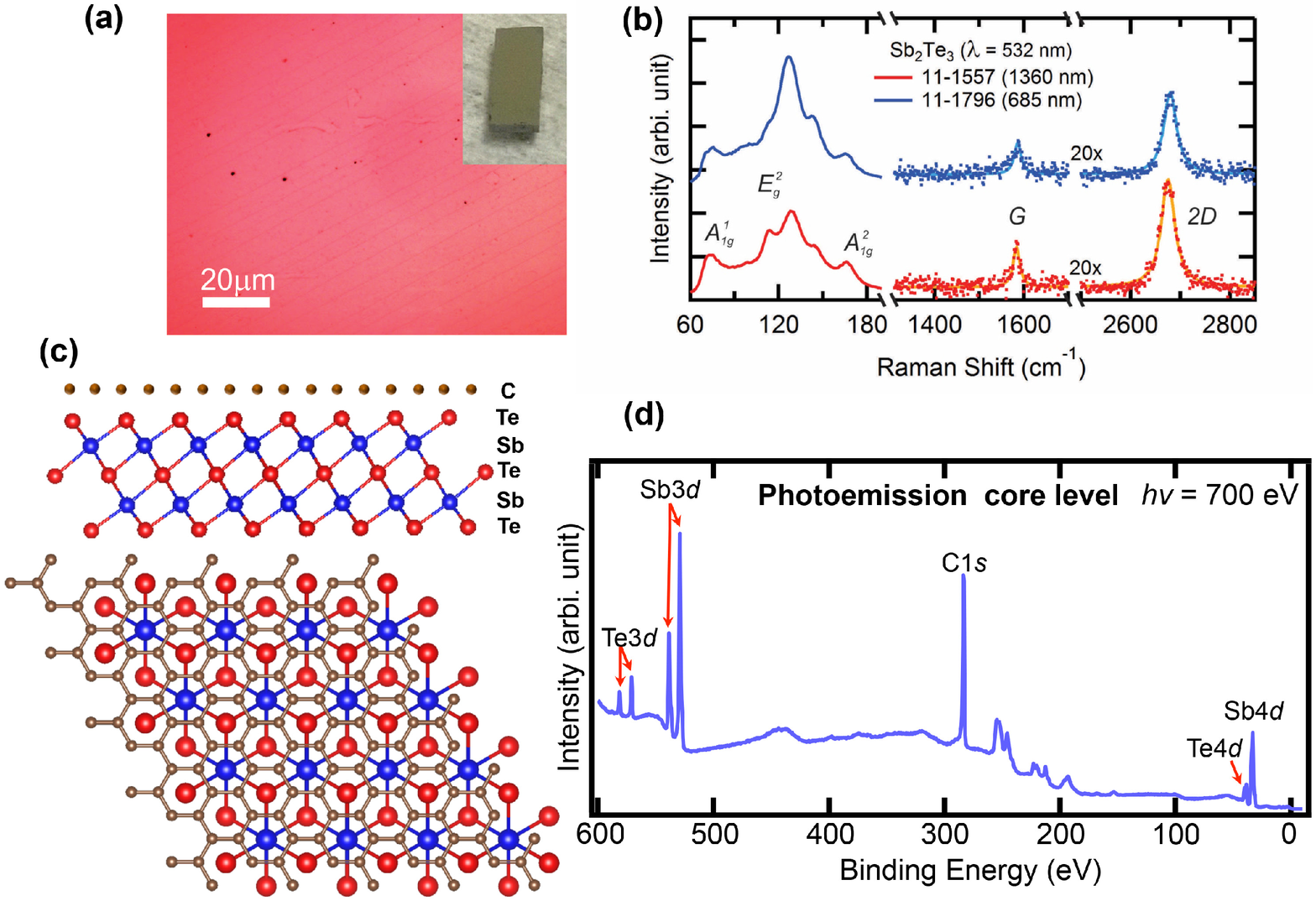}
\caption{(a) Optical microscopy image of transferred CVD graphene on Sb$_2$Te$_3$ (1360 nm thick) grown on an undoped GaAs substrate. Inset shows photograph of the sample used for ARPES measurement. (b) Raman spectra (excitation laser wavelength = 532 nm) of graphene/Sb$_2$Te$_3$ heterostructures with two different Sb$_2$Te$_3$ thin films used (thickness = 1360 nm and 685 nm). High frequency Raman spectra (1300$-$2900 cm$^{-1}$) are amplified by 20 times for clarity. (c) Schematic of lattice structure of graphene-Sb$_2$Te$_3$ heterostructure with perfect lattice match. (d) Photoemission core-level spectrum showing the chemical composition of the sample.} 
\end{figure}

\newpage

\begin{figure}
\centering
\includegraphics[width=16cm]{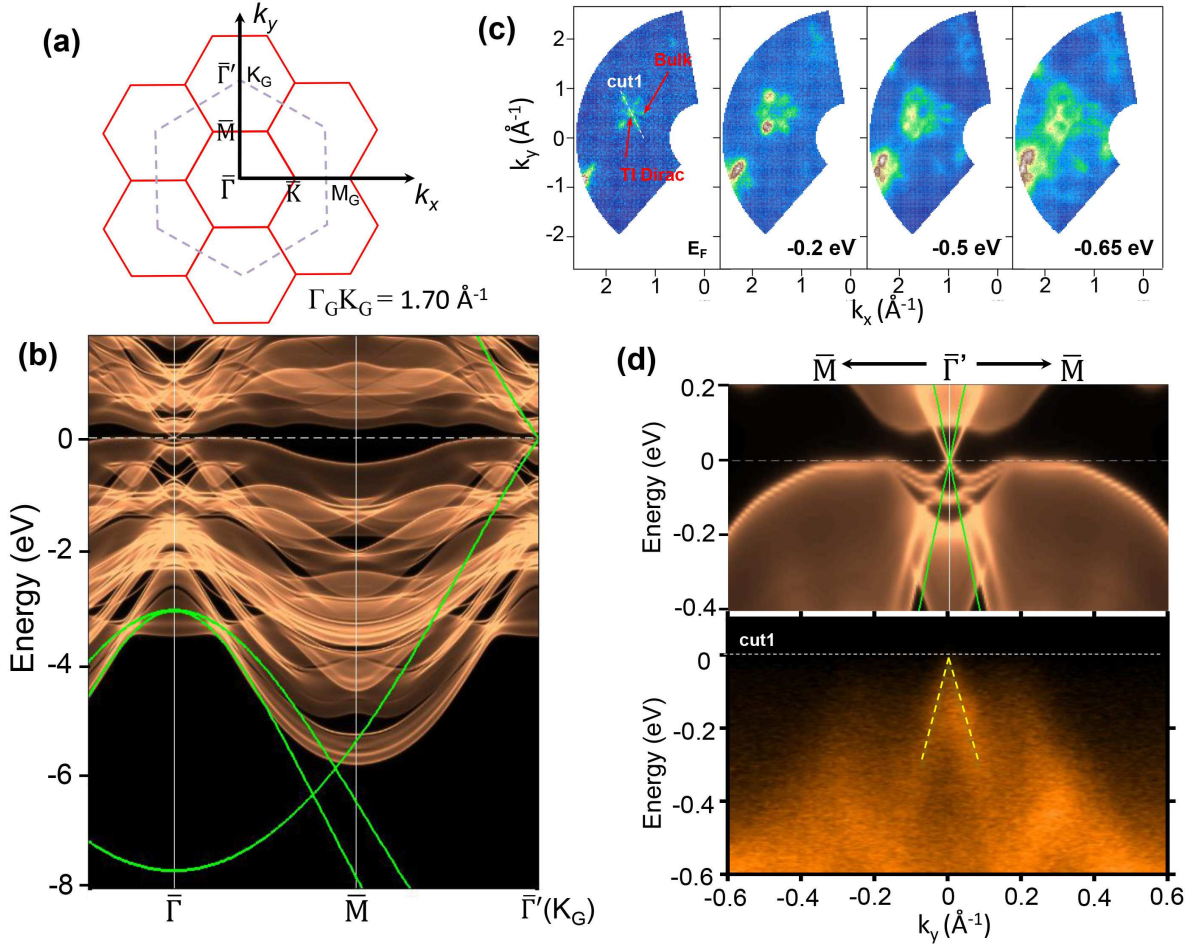}
\caption{(a) Surface Brillouin Zone of Sb$_2$Te$_3$ (red solid lines) and 1st Brillouin zone of graphene (gray dashed line), taking $\overline{\Gamma}$-$\overline{K}$ of Sb$_2$Te$_3$ as $k_x$ direction. (b) First-principles band structure of Sb$_2$Te$_3$ superimposed with that of graphene (green solid lines). (c) ARPES mapping of MOCVD-grown Sb$_2$Te$_3$ films at different binding energies. The photon energy is 100 eV. The topological surface state and bulk band lobes are marked by arrows  (d) Calculated band structure of Sb$_2$Te$_3$ (top panel, the green lines show the calculated graphene $\pi$ band for comparison purposes) and ARPES spectral cut (bottom panel, along  the dotted line in (c)) along $\bar{M}$-$\bar{\Gamma}$-$\bar{M}$ direction.} 
\end{figure}

\newpage

\begin{figure}
\centering
\includegraphics[width=16cm]{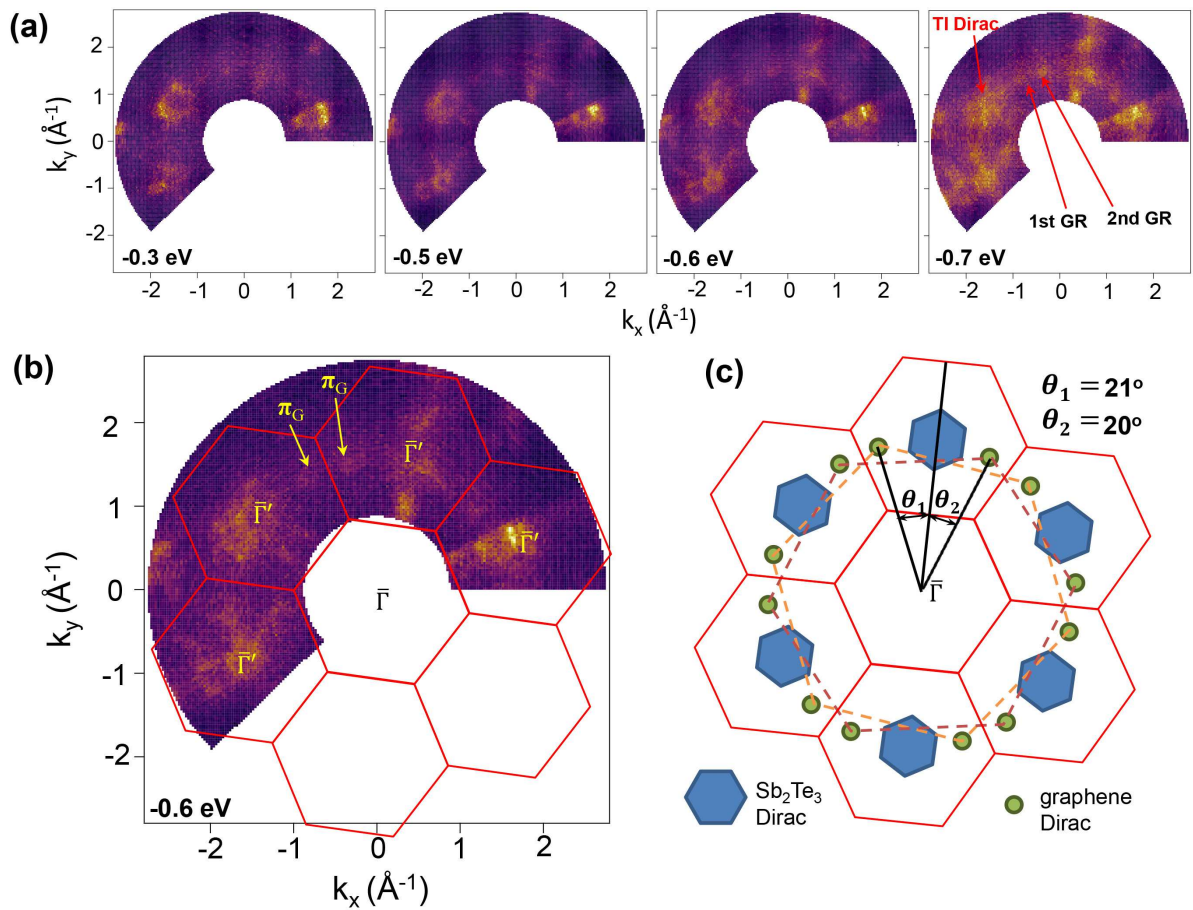}
\caption{(a) ARPES iso-energy contours of CVD graphene/Sb$_2$Te$_3$ samples at different binding energies, showing topological surface states of Sb$_2$Te$_3$ and Dirac $\pi$ bands from two graphene domains. (b) Zoom-in ARPES mapping with binding energy equal to 0.6 eV stacked with the surface Brillouin zone of Sb$_2$Te$_3$. The location of graphene $\pi$ bands are marked by arrows. (c) Schematic showing the angle mismatch between the Sb$_2$Te$_3$ film and two graphene domains. The red solid lines represent the surface BZ of Sb$_2$Te$_3$ and dashed lines BZ of two graphene domains. The blue hexagons schematically mark the position of  Sb$_2$Te$_3$ surface states and green circles the position of graphene Dirac $\pi$ bands.}
\end{figure}

\newpage

\begin{figure}
\centering
\includegraphics[width=16cm]{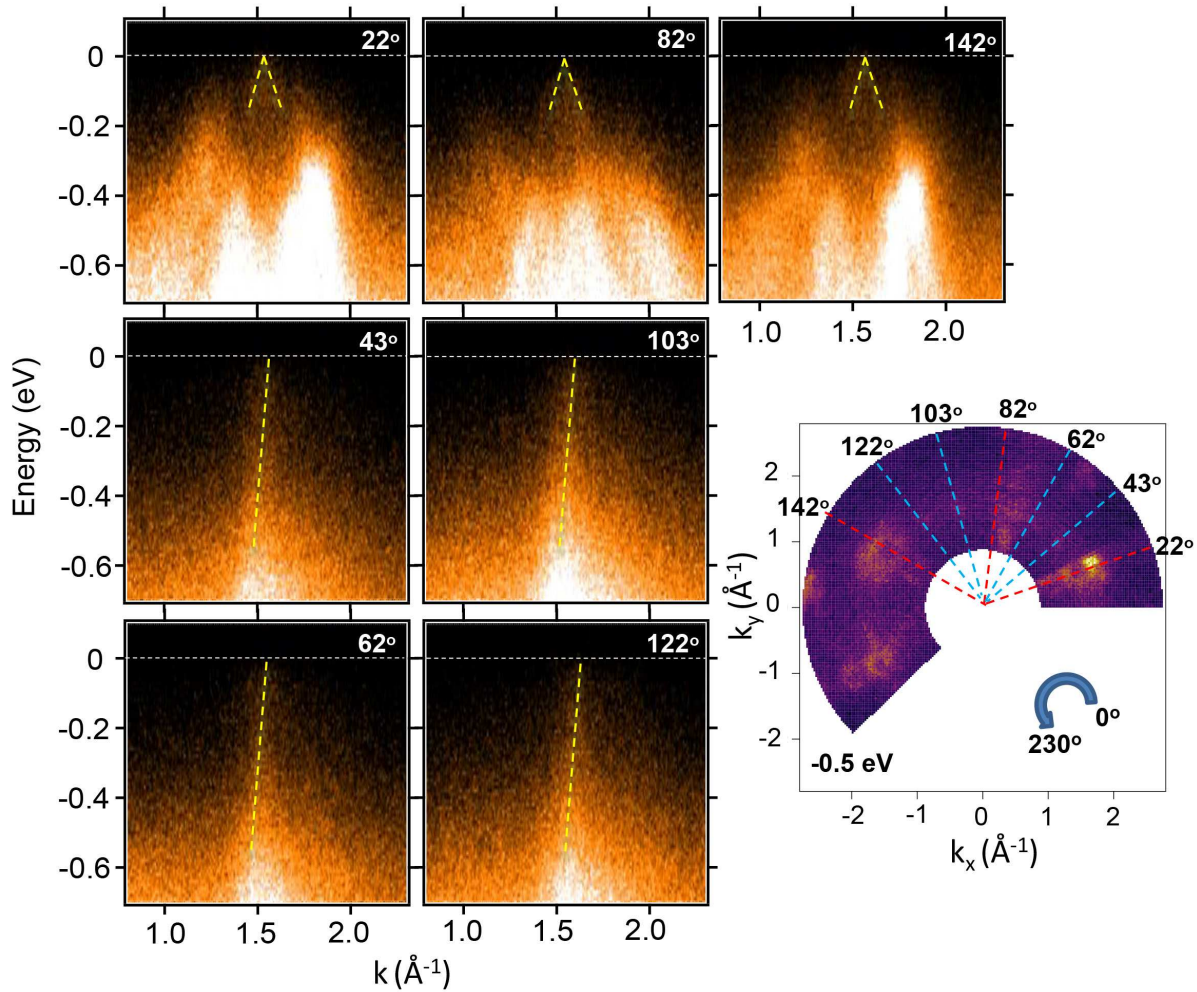}
\caption{ARPES spectral cuts at different azimuthal angles showing the coexistence of topological surface Dirac fermions of Sb$_2$Te$_3$ and Dirac $\pi$ bands of graphene. The dashed lines serve as a guide to the eye, showing the dispersion of TI Dirac surface states and graphene $\pi$ bands.}
\end{figure}

\end{document}